# Studies on Magnetic-field induced first-order transitions

P. Chaddah

UGC-DAE Consortium for Scientific Research,
University Campus, Khandwa Road, Indore 452017

*Abstract*

*We shall discuss magnetization and transport measurements in materials exhibiting a broad first-order transition. The phase transitions would be caused by varying magnetic field as well as by varying temperature, and we concentrate on ferro – to antiferro – magnetic transitions in magnetic materials. We distinguish between metastable supercooled phases and metastable glassy phase.*

## INTRODUCTION

First order phase transitions (FOPT) that can be caused by either varying temperature or by varying H were extensively studied in the context of vortex matter phase transitions [1]. Such phase transitions occur in magnetocaloric materials, colossal magnetoresistance materials, magnetic shape-memory alloys etc., and are also believed to be the cause for the functional properties of these materials [2]. In the absence of disorder, the FOPT occurs at a sharply defined line in the 2-control variable (H, T) space. Many of these functional magnetic materials are multi-component systems whose properties become more interesting under substitution. Such substitutions are an intrinsic source of frozen disorder. Early theoretical arguments of Imry and Wortis [3] showed that such samples would show a disorder-broadened transition, with a spatial distribution of the $(H_C, T_C)$ line across the sample.

Phase transitions observed in vortex matter [1] are broadened both by intrinsic disorder and by pinning giving rise to a variation of the local field across the sample. It was realized that broad 1st order transitions would not show jump discontinuities in physical properties, and Clausius-Clapeyron relation cannot be invoked for identifying the order of the transition [1].

Supercooling and superheating across a first order transition (FOPT) yield metastable states, resulting in hysteresis. But hysteresis and metastability, by themselves, are arguable signatures of a FOPT. Kinetic hysteresis is also seen when molecules in amorphous solids exhibit arrested kinetics, and equilibrium in this glassy state can not be reached over experimental time scales. Supercooled and 'glassy' states are both metastable states associated with an underlying FOPT. We shall briefly outline some features of hysteresis associated with supercooling, as these were used in establishing the nature of a 1st order transition in vortex matter. We shall then look at an interplay between these two phenomenon across disorder-broadened 1st order transitions, which have been under active study because these show phase coexistence.

## SUPERCOOLING

Following the standard treatment of supercooling across a first-order transition we considered the case when both T and density are varied to cross the phase boundary [4]. The spinodal or limit for supercooling T* is the limit below above which the supercooled state sits in a local minimum of free energy and is metastable; below this it is unstable and must convert to the ordered state unless kinetics prevents this. A supercooled state is not unstable but sits in a local minimum of the free energy. Similarly, a superheated state is metastable upto a limiting temperature T**. The extent of hysteresis observed, in a constant pressure phase transition, is dictated by the window {T** - T*}. We obtained the general result [4] that this window must increase(decrease) with increasing density for transitions in which $T_C$ falls(rises) with increasing density. We also showed that the limit to which one can supercool will be farther if no changes in density (of vortex matter) are caused during the supercooling process; cooling in constant magnetic field would allow deeper supercooling [4]. We also showed that oscillations in vortex density, caused by oscillations in applied H, would transform the metastable supercooled state into the stable equilibrium state. All these conclusions constituted a verifiable result, and were used for establishing that the transition at the peak-effect in $CeRu_2$ is a $1^{st}$ order transition [5].

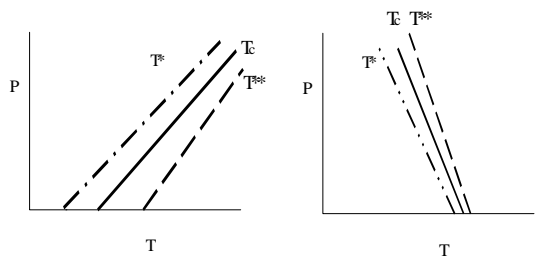

Fig. 1 : The hysteresis window encompasses smaller T-range with rising density if $T_C$ rises with rising density, and larger T-window if it falls.

## GLASS FORMATION OR KINETIC ARREST

Other than the general definition that "glass is a noncrystalline solid material that yields broad nearly featureless diffraction pattern", there exists another widely acceptable picture of glass as a liquid where the atomic or molecular motions *or kinetics are arrested*. Within this latter dynamical framework, "glass is time held still" [6]. It appears reasonable that kinetic arrest, on experimental time scale, would occur below some $T_k$. In case glass is formed in a state that is metastable (in terms of free energy) then glass formation can be caused by slow cooling, as in O-terphenyl. If the state in which glass is formed is unstable, then rapid quenching is essential so that kinetics is frozen *before* it can effect a structural change, as in metglasses. In our picture, what matters is whether $T_k < T^*$ (metglass) or $T_k > T^*$ (O-terphenyl). By studying a material in which $T_C$ varies sharply with H, one can hope to go from a metglass-like situation to an O-terphenyl-like situation. This potential situation exists in vortex matter phase transitions, as also in many magnetic materials undergoing a metamagnetic transition. This was recognized by Manekar et al [7] in the case of doped $CeFe_2$, and they also speculated that this could also be happening for the charge-ordering transition in some manganites. The schematic in fig 2 shows a situation in which the low-T ground state is antiferromagnetic (AFM) and the high-T equilibrium state is ferromagnetic (FM). This is the situation encountered by Manekar et al in doped $CeFe_2$ and in $(Nd,Sr,Sm)MnO_3$. Zero-field-cooling (ZFC) process would produce an AFM state, while field-cooling (FC) in large H would result in a kinetically arrested FM state T because the kinetics is arrested at $T_k$, and the FM-to-AFM which was to occur at the lower $T^*$.

## DISORDER-BROADENING

We had started by stating that many functional magnetic materials are multi-component systems whose properties become more interesting under substitutions, which are an intrinsic source of frozen disorder. Such samples would show a disorder-broadened transition, with a spatial distribution of the $(H_C, T_C)$ line across the sample. The spinodal lines corresponding to the limit of supercooling $(H^*,T^*)$ would also be broadened into bands for samples with quenched disorder [7]. Each of these bands corresponds to a quasicontinuum of lines; each line represents a region of the disordered sample. Similarly, the disordered system would have a $(H_K,T_K)$ band formed out of the quasi-continuum of $(H_K,T_K)$ lines. Each line in this $(H_K,T_K)$ band represents a local region of the sample and would have its conjugate in the $(H^*, T^*)$ band. The formation of the supercooling band results in phase coexistence, a phenomenon being reported with increasing frequency, since when T lies within the band some regions of the sample would have transformed and others would not have. The first visual realization of such a phase coexistence was provided by Soibel et al. [8] for the vortex melting transition. A similar visual realization of such a phase coexistence for an AFM to FM transition, in doped $CeFe_2$, was provided by Roy et al. [9]. An explanation for these observations is provided by the schematic in fig 3, adapted from ref [7]. The results in these two papers correspond to point C and path 2 because the two phases coexisted only over a finite range of fields, and full transformation could be seen in both H-increasing and H-decreasing cases. Similarly, Roy et al showed FC measurements for point A since the transformation was completed in both T-increasing and T-decreasing cases. More intriguing data results if one reaches point F following path 2, or point C and then L, following path 1. Specifically, following path 2 at 3K in doped $CeFe_2$, Manekar et al [7] found that even though H is reduced to zero, the starting AFM state is not fully recovered. The data is shown in fig 4 (a) and (b). To understand this, we look at fig 3 and path 2 leading to point F (which is at H=0). Part of the arrested FM state is converted to AFM, but full de-arrest does not take place. It follows from fig 3 that if the run is repeated at slightly higher T, then the arrested FM fraction will be smaller; the data at 5K confirmed this[7]. Similar, but less complete, data existed in literature on $(Nd,Sr)MnO_3$, and this was also explained [7]. FC data was also taken in various fields on doped $CeFe_2$, and fig 5 shows data for H=0.7 Tesla [10]. Again, following path 1 leading to points D and E in the schematic fig3, we do not recover the full AFM state obtained in the ZFC case. Points E and L will have the same frozen FM fraction, and this comes out in fig 5. Singh et al [10] took data at higher field and, as expected, the frozen FM fraction was larger. ]. Similar, data existed in literature on $(Nd,Sr,Sm)MnO_3$, and this was also explained [10]. Such part-transformations, and frozen coexisting-phases, is a manifestation of disorder-broaden transitions. It is also a strong interplay between supercooled states and kinetically-arrested (or glassy) states. Since both these are **not equilibrium or stable** states, there is an expectation of observable relaxation effects. We now address these possibilities.

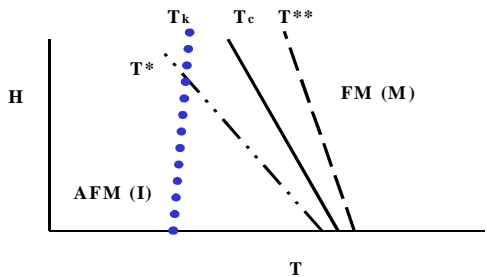

Fig 2 : The dotted curve shows the $T_k$ line which falls below $T^*$ at low H (as in a metglass), but lies above it at high H (as in O-terphenyl).

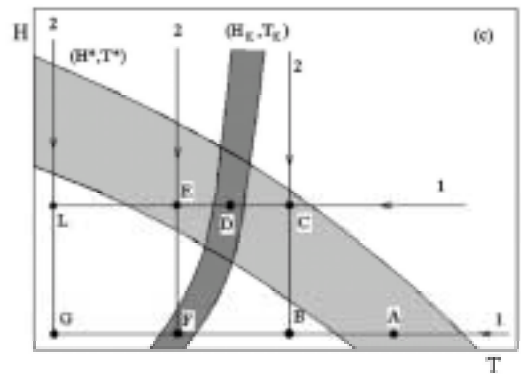

Fig 3 : Two phases will be observed at points A, C, D, E and L by paths 1 (FC); and at points C and F by paths 2 (T-fixed, lowering H), from [7]

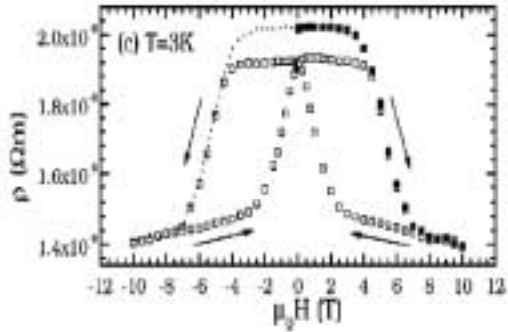

**Fig 4(a) : Isothermal reduction of field results in only part conversion of FM to AFM phase, down to H=0; from [7].**

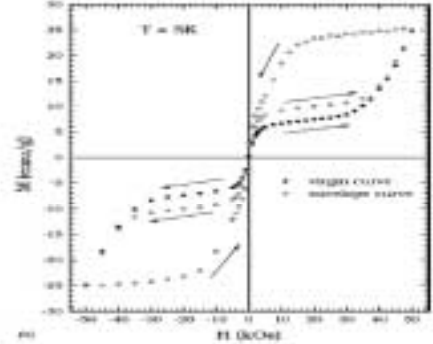

**Fig 4(b) : The virgin M-H curve lies outside the hysteresis loop, because the virgin ZFC state is AFM, whereas the remanent state at H=0 has some frozen FM [7]**

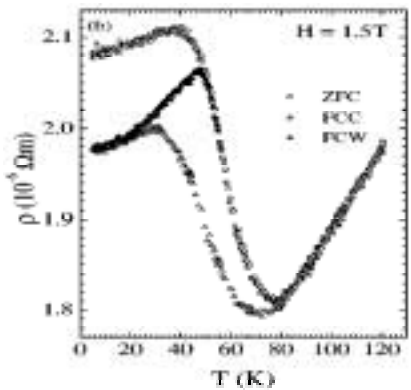

**Fig 5 : FCC results in part conversion to AFM phase; from [10]**

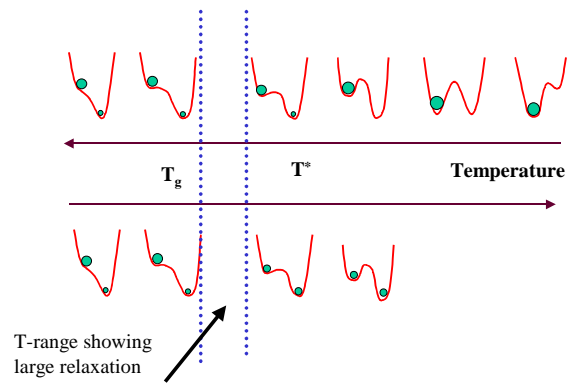

**Fig 6 : Schematic of close-lying T* and $T_k$**

**RELAXATION EFFECTS**

As discussed earlier, a supercooled state sits in a local minimum of free energy, and the barrier defining this local minimum decreases with falling T. It is thus expected that relaxation effects would **increase** as one lowers T for a supercooled state. A kinetically arrested state, on the other hand, need not be in a local minimum of free energy; its further transformation is inhibited because molecular motions are too slow. It is thus intuitive that these motions would become slower as T is lowered, and relaxation effects here would **decrease** as T is lowered. As indicated in fig 2, for cooling in some particular field value, T* could lie just above $T_k$. In this case, Relaxation rate would first rise as T is lowered towards T*, and then fall as one crosses $T_k$. This is depicted in the schematic in fig 6. Chattopadhyay et al [11] have observed this behaviour in doped $CeFe_2$. Similar behaviour has been observed by Ghivelder and Parisi in $LaPrCaMnO_3$ [12].

## DISCUSSION

I have discussed two distinct types of metastable states associated with a $1^{st}$ order phase transition. Intrinsic disorder adds newer features because the free engy landscape broadens various transition lines into bands. Features like coexisting phases, commonly observed over regions of T (or H) in many functional magnetic materials, can be understood. We have added to the complexity by adding the concept of kinetic arrest. This adds newer possibilities, which have been confirmed through measurements on the doped $CeFe_2$ system. This has helped explain intriguing observations in various manganite systems. Similar observations in the magnetocaloric material $Gd_5Ge_4$ have also been explained using these ideas [13]. These ideas have also been recently exploited to explain irreversibilities in another intermetallic $Nd_7Rh_3$ [14].

## ACKNOWLEDGEMENTS

I am grateful to the many collaborators with whom these problems have been addressed, some of whose work was not covered in this brief talk. I particularly thank Dr S B Roy for a long-standing collaboration, and Dr Alok Banerjee for discussions.